\documentclass[12pt,preprint]{aastex}

\shorttitle{Reddening of Red Supergiants}
\shortauthors{Massey et al.}
\begin{document}

\title{The Reddening of Red Supergiants: \\ When Smoke Gets in Your Eyes}

\author{Philip Massey\altaffilmark{1,2}, 
        Bertrand Plez\altaffilmark{3},
        Emily M. Levesque\altaffilmark{4}, 
        K. A. G. Olsen\altaffilmark{5},
        Geoffrey C. Clayton\altaffilmark{6},
        Eric Josselin\altaffilmark{3}}

\altaffiltext{1}{Lowell Observatory, 1400 W. Mars Hill Rd., Flagstaff,
AZ 86001.}

\altaffiltext{2}{Visiting Observer, Kitt Peak National Observatory,
National Optical Astronomy Observatory, which is operated by the
Association of Universities for Research in Astronomy, Inc., under
cooperative agreement with the National Science Foundation.}

\altaffiltext{3}{GRAAL, CNRS, Universit\'{e} de Montpellier II, 34095
Montpellier Cedex 05, France.}

\altaffiltext{4}{Massachusetts Institute of Technology, 77
Massachusetts Avenue, Cambridge, MA 02139.}

\altaffiltext{5}{Cerro Tololo Inter-American Observatory, National
Optical Astronomy Observatory, Casilla 603, La Serena, Chile.}

\altaffiltext{6}{Department of Physics and Astronomy, Louisiana
State University, Baton Rouge, LA 70803.}

\begin{abstract}

Deriving the physical properties of red supergiants (RSGs) depends upon
accurate corrections for reddening by dust.  We use our recent modeling
of the optical spectra of RSGs to address this topic.  
First, we find
that previous broad-band studies have underestimated the correction for
extinction in the visible, and hence the luminosities (if derived from
$V$); the shift in the effective wavelengths of the standard $B$ and
$V$ bandpasses necessitates using an {\it effective} value of the ratio
$R'_V=4.2$ to correct broad-band
photometry of RSGs if $R_V=3.1$ for early-type stars viewed through the
same dust, where we have assumed the standard reddening law of Cardelli,
Clayton, \& Mathis (1989).  Use of the Fitzpatrick (1999) reddening law would lead
to $R'_V=3.8$, as well as slightly lower values of extinction derived
from spectrophotometry, but results in slightly poorer fits.
Second, we find that a significant fraction of RSGs in
Galactic OB associations and clusters show up to several magnitudes of
excess visual extinction compared to OB stars in the same regions; we
argue that this is likely due to circumstellar dust around the RSGs.
We also show that the RSG dust production rate (as indicated by the
12-$\mu$m excess) is well-correlated with bolometric luminosity,
contrary to what has been found by earlier studies.  The stars with the
highest amount of extra visual extinction also show significant near-UV (NUV)
excesses compared to the stellar models reddened by the standard
reddening law.  This NUV excess
is likely due to scattering of the star's light by the dust and/or a larger average
grain size than that typical of grains found in the diffuse interstellar medium.
Similar excesses have been attributed to circumstellar dust around R Coronae
Borealis stars.  Finally, we estimate that the RSGs 
contribute dust grains at the rate of
$3 \times 10^{-8}  M_\odot$ yr$^{-1}$ kpc$^{-2}$ in
the solar neighborhood, comparable to what we estimate for late-type
WCs, $1 \times 10^{-7} M_\odot$ yr$^{-1}$ kpc$^{-2}$.  
In the solar neighborhood this represents
only a few percent of the dust production (which is dominated
by low-mass AGBs), but we note that in low-metallicity starbursts,
dust production by RSGs would likely dominate over other sources.

\end{abstract}


\section{Introduction}
\label{Sec-intro}

Red supergiants (RSGs) are the evolved, He-burning descendents of
moderately massive ($\leq 40 M_\odot$) O and B stars.  We have recently
used the new generation of MARCS stellar atmosphere models 
(Gustafsson et al.\ 1975, 2003; Plez et al.\ 1992)
to fit optical spectrophotometry of 74 Milky Way
RSGs, from which we derived a new effective temperature scale for RSGs
of Galactic metallicity
(Levesque et al.\ 2005, hereafter Paper I).  A subsample of
62 of these stars belong to OB associations with known distances, which
allowed us to determine other physical properties, such as bolometric
luminosity and stellar radii, for comparison with those predicted by
stellar evolutionary models.  We found excellent agreement, thus
removing a major discrepancy between ``observation" and theory for
massive star evolution (see Massey 2003).

However, that study underscored the difficulty in correcting for
reddening due to dust for these objects; our work hinted at several
previously neglected problems, which we consider more fully here.
First (Sec.~\ref{Sec-Rv}), we find that previous efforts involving
broad-band filter photometry have led to systematically underestimating
the visual extinctions, and thus to underestimating the stellar
luminosities.  Although we successfully avoided that problem in Paper~I
by using spectrophotometry to derive reddenings, our work revealed that
many of these stars suffer extinction beyond that of their neighboring
O and B stars.  We show here (Sec.~\ref{Sec-Av}) that this is likely
caused by circumstellar dust from these ``smoky" stars.  Although
future studies will attempt to derive the reddening law for this dust,
 it is already clear that the circumstellar dust results in a near-UV excess
compared to the stellar models reddened with the ``standard" reddening law
(Sec.~\ref{Sec-NUV}).  Finally,
we use the results of Paper I and the current study to estimate the
fraction of dust deposited in the diffuse interstellar medium compared
to that of other sources in the Galaxy (Sec.~\ref{Sec-cosmic}).

\section{The Effective $R_V$ for Broad-band Photometry of RSGs}
\label{Sec-Rv}

A knowledge of the ratio of total to selective extinction $R_V\equiv
A_V/E(B-V)$ is needed in order to derive physical parameters of
reddened stars.  Using {\it early-type} stars with moderate reddenings
in the Milky Way, Sneden et al.\ (1978) and others found $R_V$ to be
typically 3.1, although it is now well understood that this value is
not universal in nature, but represents the average sightline through 
the diffuse interstellar medium (Valencic et al.\ 2004).
Values of $R_V\sim 5$ and even higher have been found for stars in
dense molecular clouds; see, for example, Cardelli \& Wallerstein (1989)
and Cardelli et al.\ (1989, hereafter CCM89).   These {\it physical}
variations of $R_V$ are due to differences in the line-of-sight
environment, such as the grain size distribution.

However, aside from such real differences in the dust properties, one
must employ a larger {\it effective} value of $R_V$ (which we will
denote as $R'_V$) when correcting broad-band photometry of objects
whose spectral energy distributions differ from that of lightly- to
moderately-reddened O stars.  This is simply due to the shift of the
effective wavelength of the filters compared to that obtained with
early-type stars.  A very red star shifts the effective wavelengths of
the $B$ and $V$ filters to longer wavelengths, where extinction is
less, making both $A_B$ and $A_V$ smaller for a given amount of dust.
The net effect of lowering $A_B$ and $A_V$ is to {\em increase}
$R'_V$, as the decrease in the denominator of the ratio outweighs the
decrease of the numerator.  McCall (2004) has recently emphasized the
importance of considering the spectral energy distribution (SED) of the
source when correcting for extinction in galaxy photometry, as their
SEDs do not, after all, resemble that of the OB stars for which the
extinction laws were derived.

\begin{figure}
\plotone{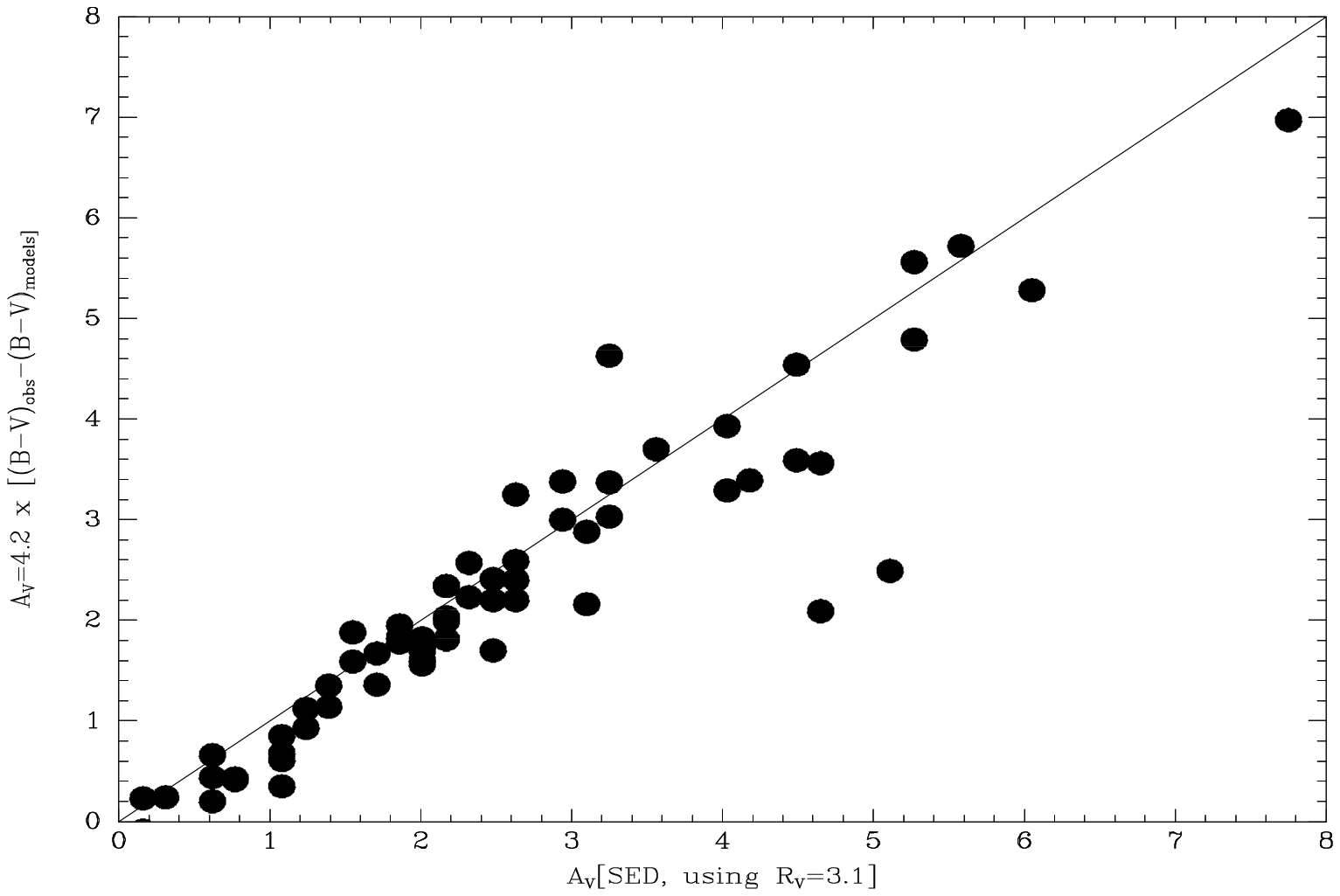}
\caption{\label{fig:Avs} Comparison of visual extinction $A_V$.
The extinction derived from broad-band photometry and $R_V=4.2$
is compared with that
derived from the spectral energy distribution and $R_V=3.1$.
}
\end{figure}

We avoided this problem in Paper~I by using spectrophotometry
of the stars to measure the color excess with regard the MARCS
models, employing the CCM89 reddening law.  Since neither the
{\it B} nor {\it V} filters were actually used in our color excesses,
conversion to $A_V$ was straight-forward, although it did assume that
the reddenings had a ``typical" $R_V=3.1$ value.  For broad-band
photometry of RSGs, a value of $R'_V=3.6$ has usually been adopted (Lee
1970, Humphreys 1978), and is consistent with the scaling of $R'_V$
with color proposed by Schmidt-Kaler (1982).  However, our work in
Paper~I has forced us to conclude that this value is not very accurate,
as it produced $A_V$ values that were systematically smaller than
extinction values derived from our spectrophotometry.  We have derived
an improved value for the effective $R'_V$ of RSGs by convolving the
MARCS models with the standard $B$ and $V$ band-passes (Bessell 1990).
We used the CCM89 reddening law with $R_V=3.1$ to add various amounts
of reddening to the model spectra; we then computed the resulting
change in the flux in the $V$ filter to find $A_V$, and the differences
in the relative fluxes in the $B$ and $V$ filters (compared to the
no-reddening case) in order to determine $E(B-V)$.  The effective value
of $R'_V=A_V/E(B-V)$ then follows directly.
We found that there is a significant dependence
upon the assumed surface gravity $\log g$ in the sense that stars with
higher $\log g$ (0.5) produce lower $R_V$ values; the value $R'_V=3.6$
usually used for RSGs is actually more appropriate to the surface
gravities of red dwarfs or giants.  There is very little dependence of
$R'_V$ on 
effective temperature in this regime.  This is not surprising, as it is
well known that $B-V$ is primarily sensitive to surface gravity rather
than temperature for RSGs; see discussion in Massey (1998).  Stars with
higher reddenings require a slightly higher $R'_V$ value.  In summary,
for stars with $\log g$ between -0.5 and 0.5, $T_{\rm eff}$ between
3400 and 4200 K, and $E(B-V) \leq 3$, we find
$$R'_V=4.1+0.1E(B-V)-0.2\log g.$$ For a ``modestly" reddened RSG, a
value of $R'_V=4.2$ is thus appropriate.   As
shown in Fig.~\ref{fig:Avs}, this value for $R'_V$ now brings the
extinction derived from broad-band photometry into accord with that
derived from spectrophotometry.

\begin{figure}
\epsscale{0.42}
\plotone{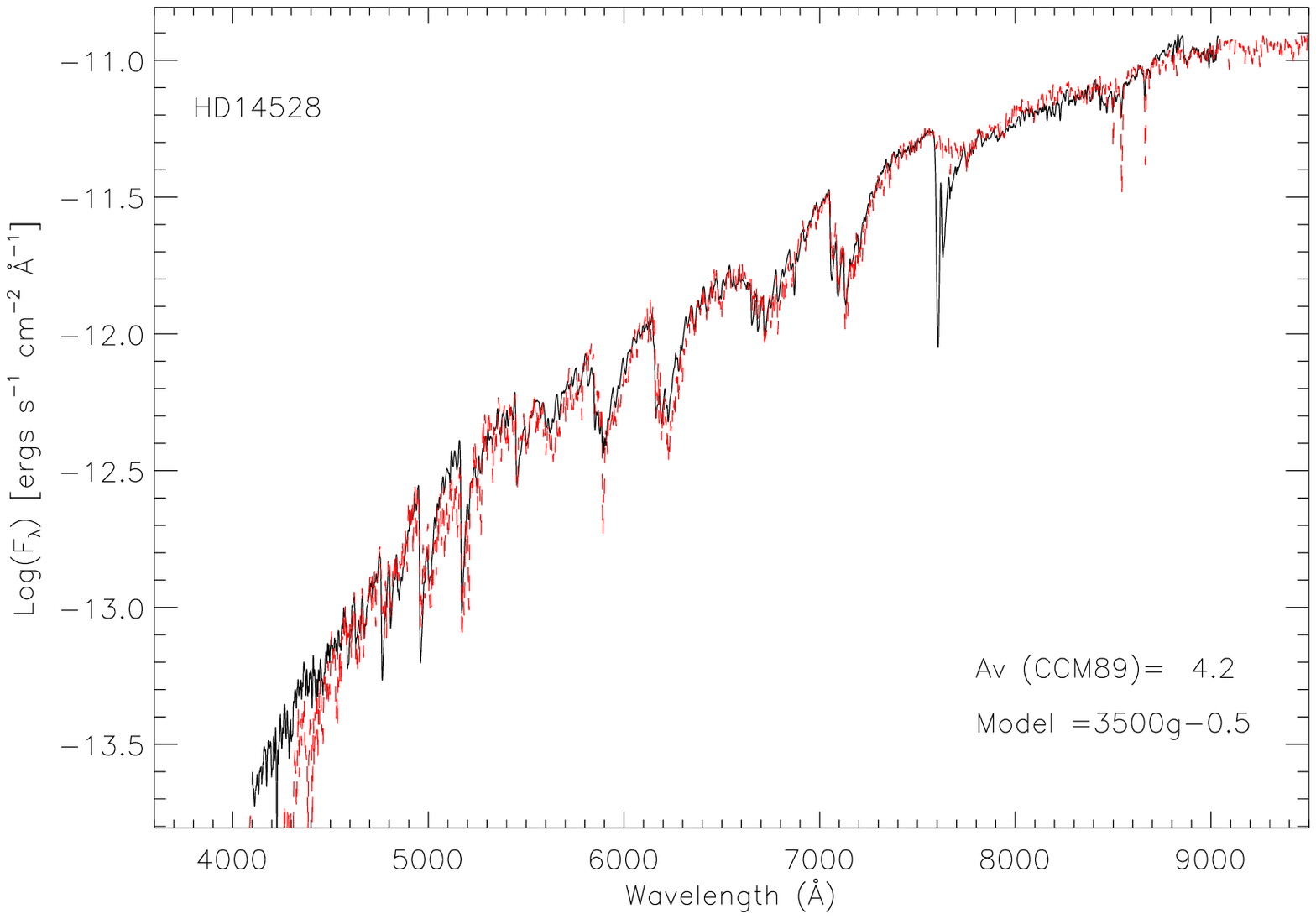}
\plotone{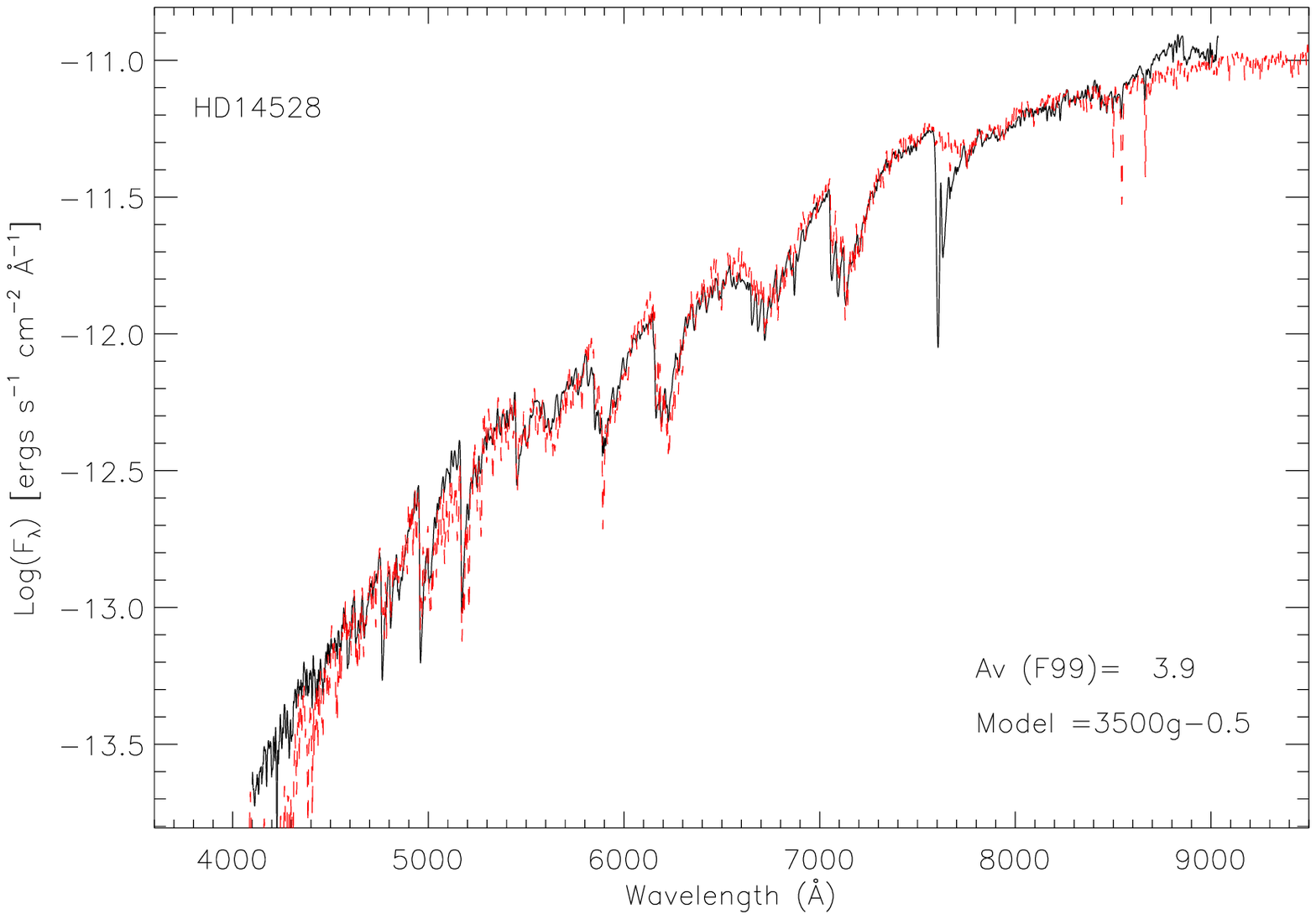}
\plotone{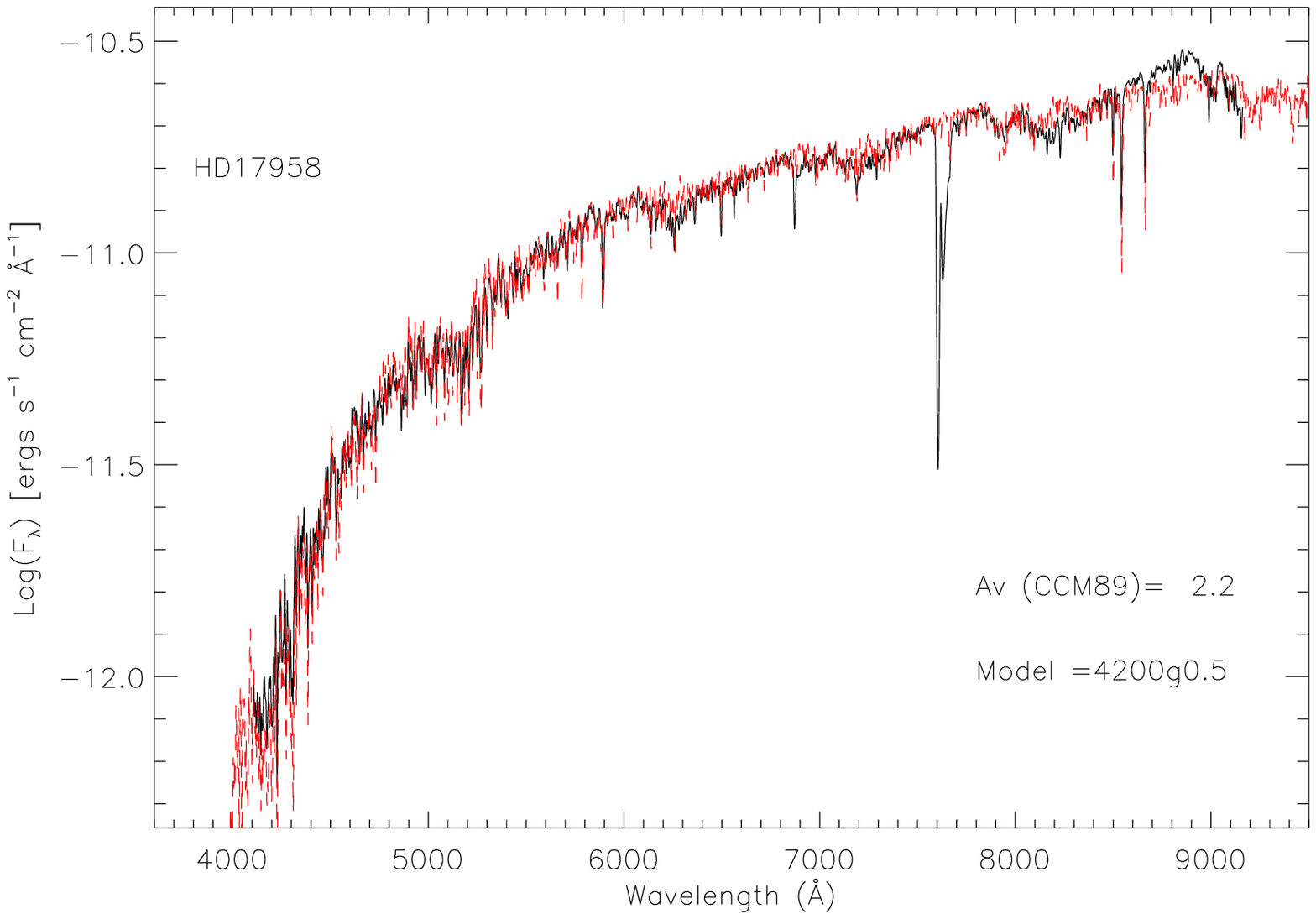}
\plotone{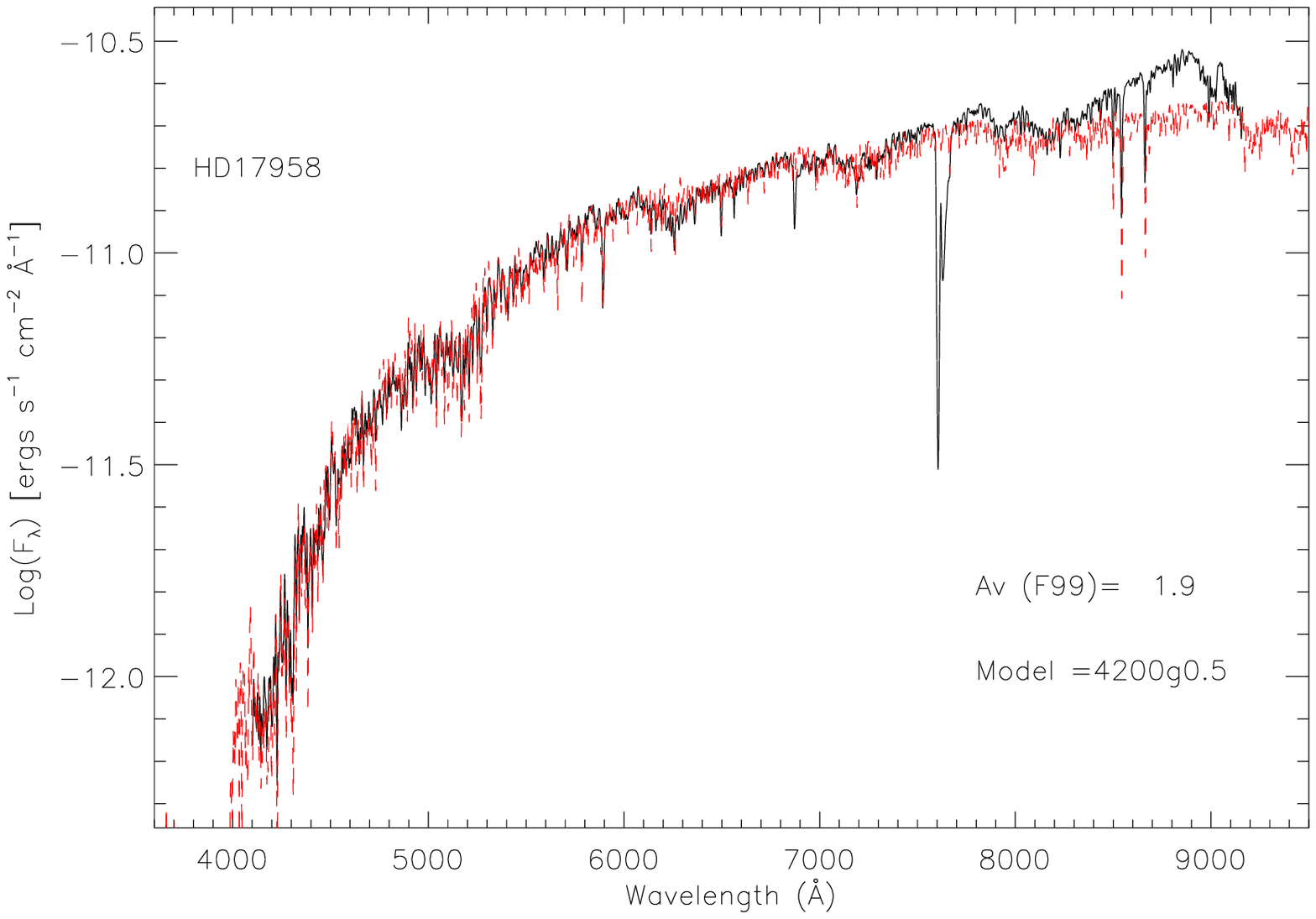}
\caption{\label{fig:laws} Comparison of the fits obtained
CCM89 and Fitpztrick (1999, hereafter F99)  reddening laws.
In the top two panels, we
compare the fits for HD 14528 (M4.5 I) using
the MARCS 3500g-0.5 model reddened by (a) CCM89 with $A_V=4.2$
and (b)  F99  with $A_V=3.9$.
In the bottom two panels, we make a similar comparison for HD 17958 (K2 I)
by reddening the MARCS 4200g0.5 model using
(c) CCM89 with $A_V$=2.2, and (d) F99 with $A_V=1.9$.  Similar
fits are obtained using  CCM89 and F99, although the agreement is 
slightly better in the far red
with the CCM89 law. The strong, unfit feature at 7600\AA\ is the telluric A-band, and
the turndown at 8900\AA\ is also due to telluric absorption.
}
\end{figure}

We were initially concerned
about such a large value for $R'_V$, but found that we derive a very similar number
using the Kurucz (1992) ATLAS9 models in place of the MARCS models. 
We also find a large $R'_V$ value when we instead employ the Fitzpatrick (1999)
reddening law with the MARCS models:
$R'_V=3.8 -0.2\log g,$ with a negligible correction for color excess 
$E(B-V) \leq 3$.  The Fitzpatrick (1999) law has slightly lower extinction
than CCM89 on the short wavelength side of the $B$ filter, and on the long 
wavelength
side of the $V$ filter; the latter is primarily responsible for the difference in
the derived
$R'_V$.
Fitzpatrick (1999) has emphasized the difficulties encountered 
in deriving a monochromatic law from broad-band photometry, and derived
a reddening curve that was
very similar to that of CCM89; but for stars of extreme color, such as RSGs, small differences 
matter. To obtain
satisfactory fits to our spectrophotometry using the Fitzpatrick (1999) reddening
law we found that we needed to decrease the visual extinction by about 0.3~mag
compared to that needed with the CCM89 law.  Even so, the CCM89 law (used in
Paper I) resulted in slightly better fits, as shown by the two examples in 
Fig.~\ref{fig:laws}.  To answer which reddening law is really ``right" requires
careful comparison with spectrophotometry of stars over a wide range of
properties.  Here we simply note that the 0.3~mag systematic differences obtained
between these two laws are comparable to the estimated uncertainties in Paper I.

\section{Evidence for Circumstellar Dust Extinction}
\label{Sec-Av}

\begin{figure}
\epsscale{1.0}
\plotone{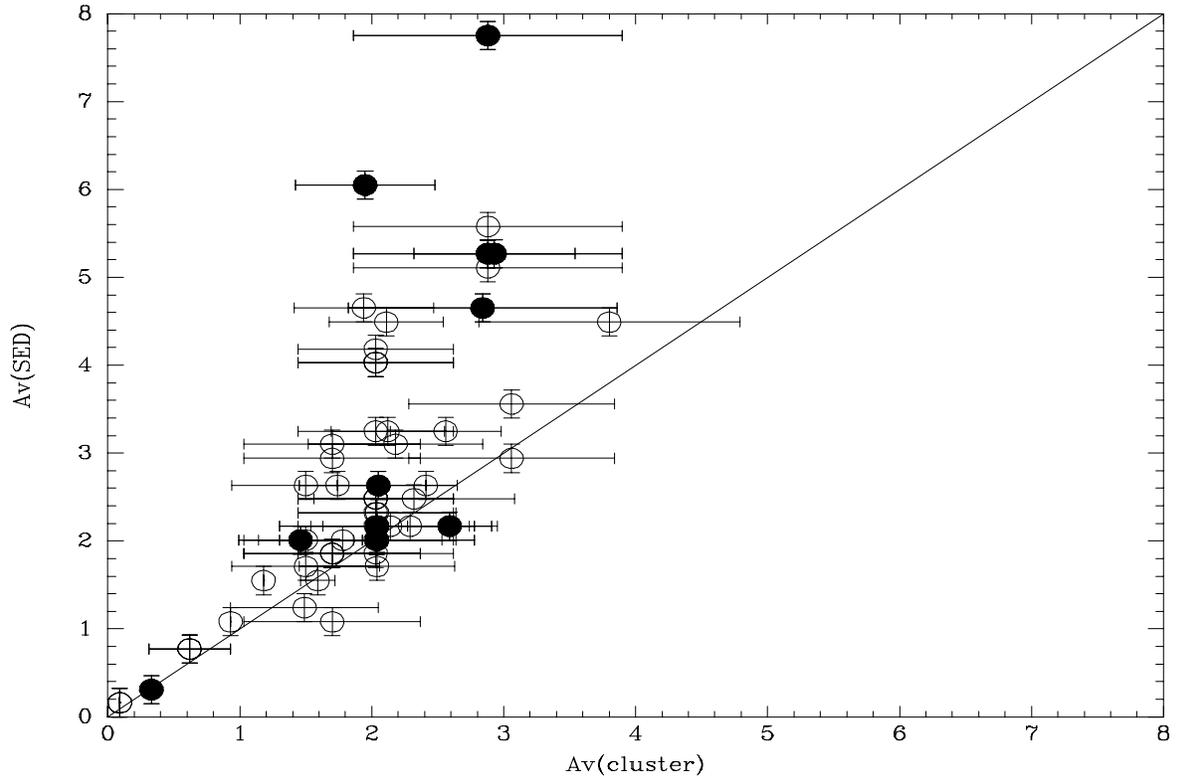}
\caption{\label{fig:Avcluster} Extinction $A_V$ of RSGs compared to
that of early-type stars in the same clusters.
The error bars denote the (1$\sigma$) spread of
extinction in each cluster and uncertainty in the $A_V$
from our fitting procedure in Paper I.  The filled circles
denote the stars for which we have new data in the near-UV.
}
\end{figure}

Although our spectrophotometry was immune to issues of bandpass shifts,
it still supposed that the extinction was due to $R_V=3.1$ reddening
that characterizes dust in the diffuse interstellar medium.  That
assumption was probably not completely valid, as circumstellar dust may
be present in some cases.  We began to consider this possibility when
we were struck
by the fact that many of the RSGs in OB associations have significantly
higher extinction than the average early-type star in the same
regions.  In Fig.~\ref{fig:Avcluster} we show the comparison between
the two.  The average $A_V$ for each cluster comes from Paper~I,
and is based upon a critical inspection of the values listed in
Humphreys (1978), eliminating stars whose spectroscopic parallaxes are
significantly (1~mag) deviant from the cluster averages.  Indeed, this
conclusion could have been made by others based purely on those data.
Of course, many of the OB associations cover a large area of the sky
and include a significant range of reddenings, and we indicate the
standard deviation by the error bars ($\pm1\sigma$) in
Fig.~\ref{fig:Avcluster}.   We find that about 40\% of our sample (22
out of 56 stars) shows evidence of extra extinction
at the $>1\sigma$ level; 14\%  (8 out of 56) shows $>3\sigma$ excess extinction.

In retrospect, this result should have been long anticipated. 
Red supergiants are known to be ``smoky" in the sense that dust 
condenses in the stellar winds.  The presence of the resulting 
circumstellar dust shells was first revealed by ground-based IR 
photometry (see, for example, Hyland et al.\ 1969), while {\it IRAS} 
two-color diagrams established that such dust shells are a common phenomenon 
for RSGs (Stencel et al.\ 1988, 1989).  This dust is thought to be 
partially responsible for driving the stellar wind via radiation 
pressure (but see the questions raised by MacGregor \& Stencel 1992).
Interferometry in the IR has demonstrated that for some RSGs (such as
VY CMa) the dust is found very close to the star itself 
(3-5 stellar radii), while in other cases it is found at greater 
distances, suggesting that the production of substantial amounts of dust 
is episodic in nature, with time scales of a few decades (Danchi et al.\ 1994).

Josselin et al.\ (2000) use the $K_o$-[12] color to determine 
dust-productions rates $\dot{M}_d$, where [12] is the magnitude
based on the IRAS 12$\mu$m flux adjusted so that a 10,000 K star would
have zero color. (Thus stars with positive $K_o$-[12] values have
some 12-$\mu$m excess, as the index has little sensitivity to effective
temperature.) One of the surprising results from that
study was that there was little or no correlation of $\dot{M}_d$ with
bolometric luminosity. From first principles one would expect that
mass-loss rates will be dependent upon luminosity, among other factors.
However, Josselin et al.\ adopted the individual spectroscopic
parallaxes listed by Humphreys (1978) as the true distances, rather than
using the average cluster values.  This is equivalent to adopting a single
absolute visual magnitude for all RSGs of a given spectral type and luminosity
class, a poor approximation since 
RSGs span a large range in masses and luminosities.
We revisit this issue here by using
the cluster distances adopted in Paper~I (based upon the early-type stars),
and by revising the $\dot{M}_d$ values using the larger and more homogeneous
$K$ dataset of the 2Mass point source catalog.  
Fig.~\ref{fig:eric}(a) shows a more convincing
correlation.  The scatter is still large, but 
in part this may be due to the inclusion
of a few unreliable data points: for instance, the two outliers on the
left are HD~160371 in M6, a cluster which contains no early-type
members (Humphreys 1978), and BD+56$^\circ$2793 (ST Cep) in Cep~OB2,
which Humphreys (1978) 
characterizes as a ``doubtful member" of its association.
In Figs.~\ref{fig:eric}(b) and (c) we show that the extra extinction
$\Delta A_V$ is correlated with $\dot{M}_d$ and with $M_{\rm bol}$.

We can also ask if this amount of extinction is {\it reasonable} given
the measured dust production rates.   Equation 3.31 from Whittet (2003) 
provides the link between the mass density of dust $\rho_{\rm d}$ 
and $A_V$ per unit path-length L:
$\rho_{d}=3.1\times 10^{-4} (A_V/L)$, where the density and length
are in MKS units.
A rigorous calculation for dust condensing
around RSGs is complicated by the fact that the dust may be decoupled from
the stellar wind gas velocities (MacGregor \& Stencel 1992) and likely occurs
episodically (Danchi et al.\ 1994), and thus the radial
distribution of dust throughout
the $10^5-10^6$ yr lifetime of the RSG phase is not easily known.  However,
{\it most} of the extinction will occur in the thin shell near where the
dust first condenses, as this is where the mass density will be the highest.
We can therefore get a crude lower-limit to the expected amount of 
extinction by assuming a thin-shell approximation: 
$A_V=\Delta R \times 3.2 \times 10^{3} \times M_d/(4\pi R^2\Delta R)$, where
$R$ is the radius (meters) where the dust condenses, 
$M_d$ is the mass (kg) of the dust,
and the shell thickness $\Delta R$ is substituted for the path length $L$,
which cancels with $\Delta R$ in the denominator.  For 10 yrs of dust production 
(a typical time scale for episodic dust formation according to Danchi et
al 1994) at a rate of
$10^{-8} M_\odot$ yr$^{-1}$, we would expect a mass of 
$2\times 10^{23}$ kg to be
deposited in this thin shell, and thus
we would have
$A_V=1.0$ mag, where we have adopted a radius $R$ of $10^4 R_\odot$
(i.e., 10 stellar radii).  Dust produced over a longer period of time,
or which condenses at a smaller radii, will increase this ``minimum reasonable
value", while a lower dust production rate, or a larger $R$, will reduce
the value.  This exercise suggests
that for the lower luminosity RSGs, $\Delta A_V$ would be a few
tenths of a mag or less (and hence not detectable), while the most extreme
RSGs might have $\Delta A_V$ values of several magnitudes---just what we find.

\section{Dust in the Near-UV}
\label{Sec-NUV}

Sufficient dust to cause several mags of visual extinction
might make its presence known in ways other than the IR excess
discussed above.  R Coronae Borealis stars, whose extreme variability
is known to be due to circumstellar dust, show a UV-excess
(Hecht et al.\ 1984, 1998).  In part, such ``fresh" dust may start
 with a distribution skewed to larger particle sizes (which would lead to less
UV and NUV extinction)
before being broken down and assimilated into the interstellar medium 
(see also Jura 1996 and Whittet 2003).  Alternatively, preferential scattering of blue light
into the beam by the parts of the unresolved circumstellar dust shell that are off-axis to the
line of sight may be enough to explain the larger observed fluxes in the UV and near-UV regions.

Indeed, as we reported in Paper I, the MARCS models reddened with an
$R_V=3.1$ CCM89 law showed this sort of NUV mismatch with the observed
spectral energy distributions for the most heavily reddened stars (which
were also the ones with the highest excess reddening).  However, we could
not completely rule out an observational explanation for the discrepancy,
given that even a small amount of red light scattered within the instrument
could contaminate the
low fluxes in the NUV without such problems showing up in the spectrophotometric
standards. For the most reddened M supergiants in our 
sample, $F_{\lambda 7000}/F_{\lambda 3500}$ is 10,000, while this ratio is near unity for standards.

On UT 2005 April 21 we obtained new data in the blue (3500-5800\AA)
on a subsample of 11 stars from Paper I using the Kitt Peak 2.1-m telescope 
with the GoldCam spectrometer.  The observational parameters are the same
as given in Table~3 of Paper~I {\it except} that a CuSO$_4$ filter was 
employed to eliminate any possibility of contamination of the NUV
region by red light.
The observational and reduction procedures were the same as given in Paper~I.
The new spectra agreed well with the old data in general, although 
turns-ups in the blue for the two reddest stars are now eliminated, suggesting that
instrument effects could indeed have been a problem with some of those data.
We therefore restrict our discussion only to the 
new data.

\begin{figure}
\epsscale{0.42}
\plotone{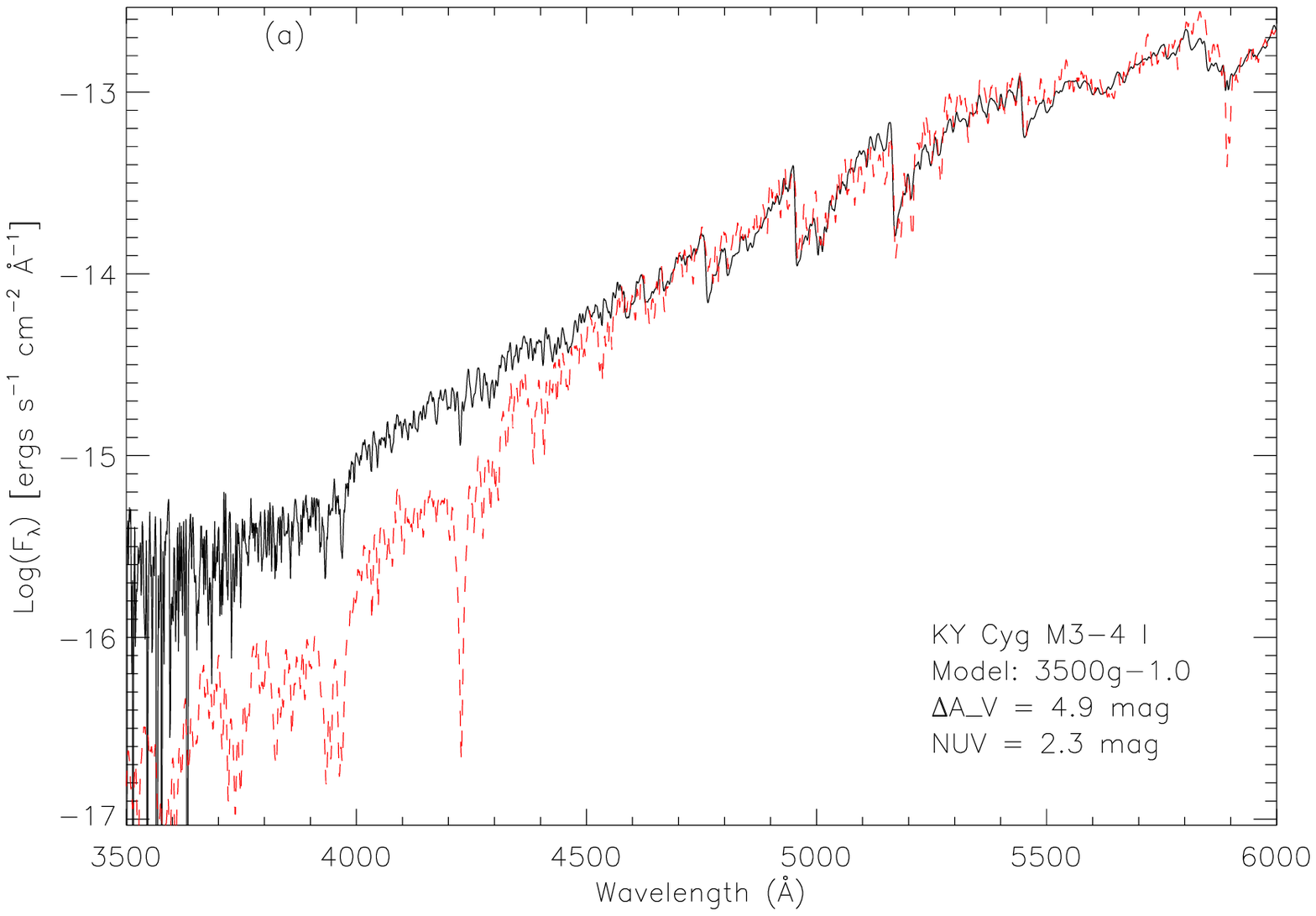}
\plotone{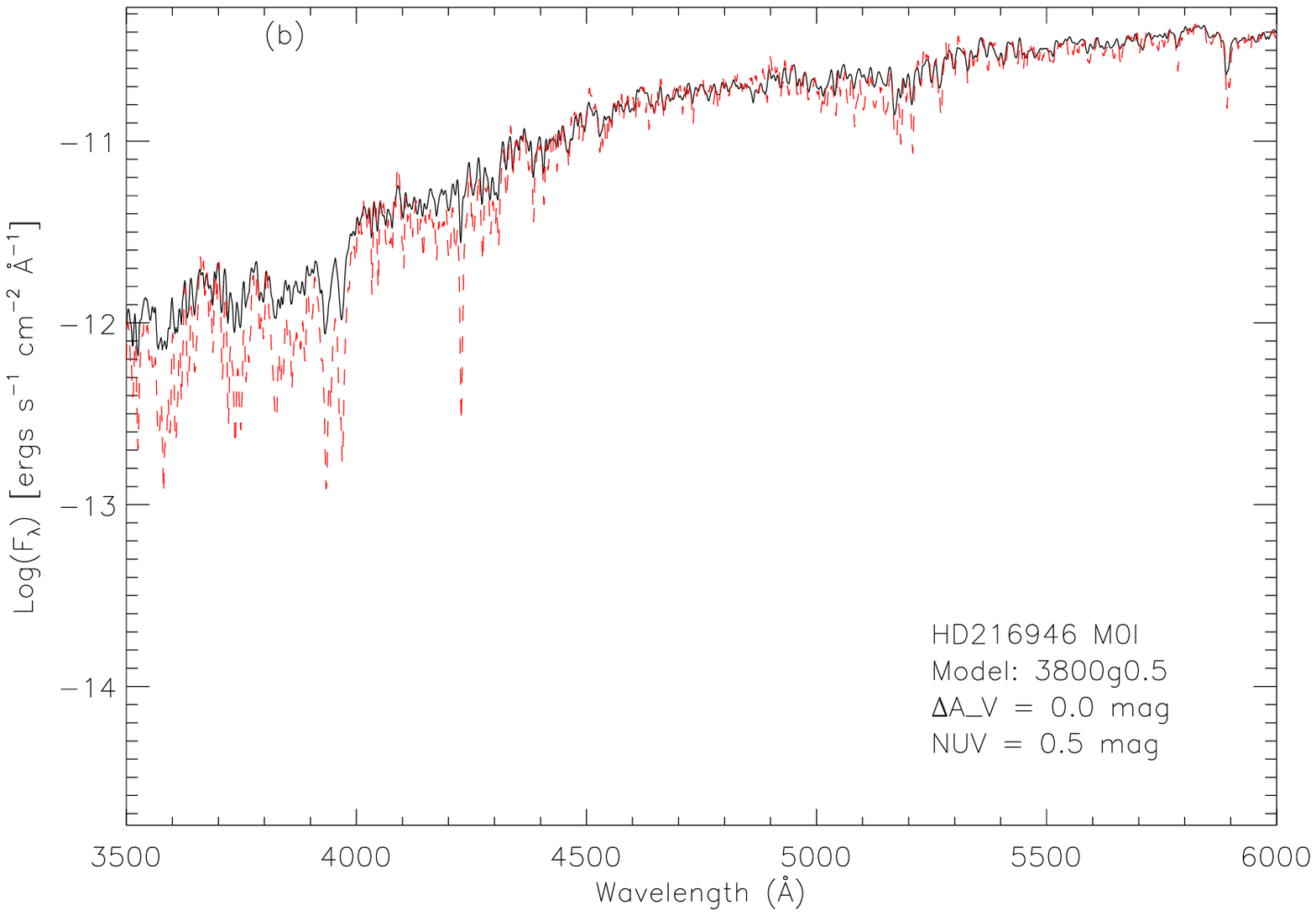}
\plotone{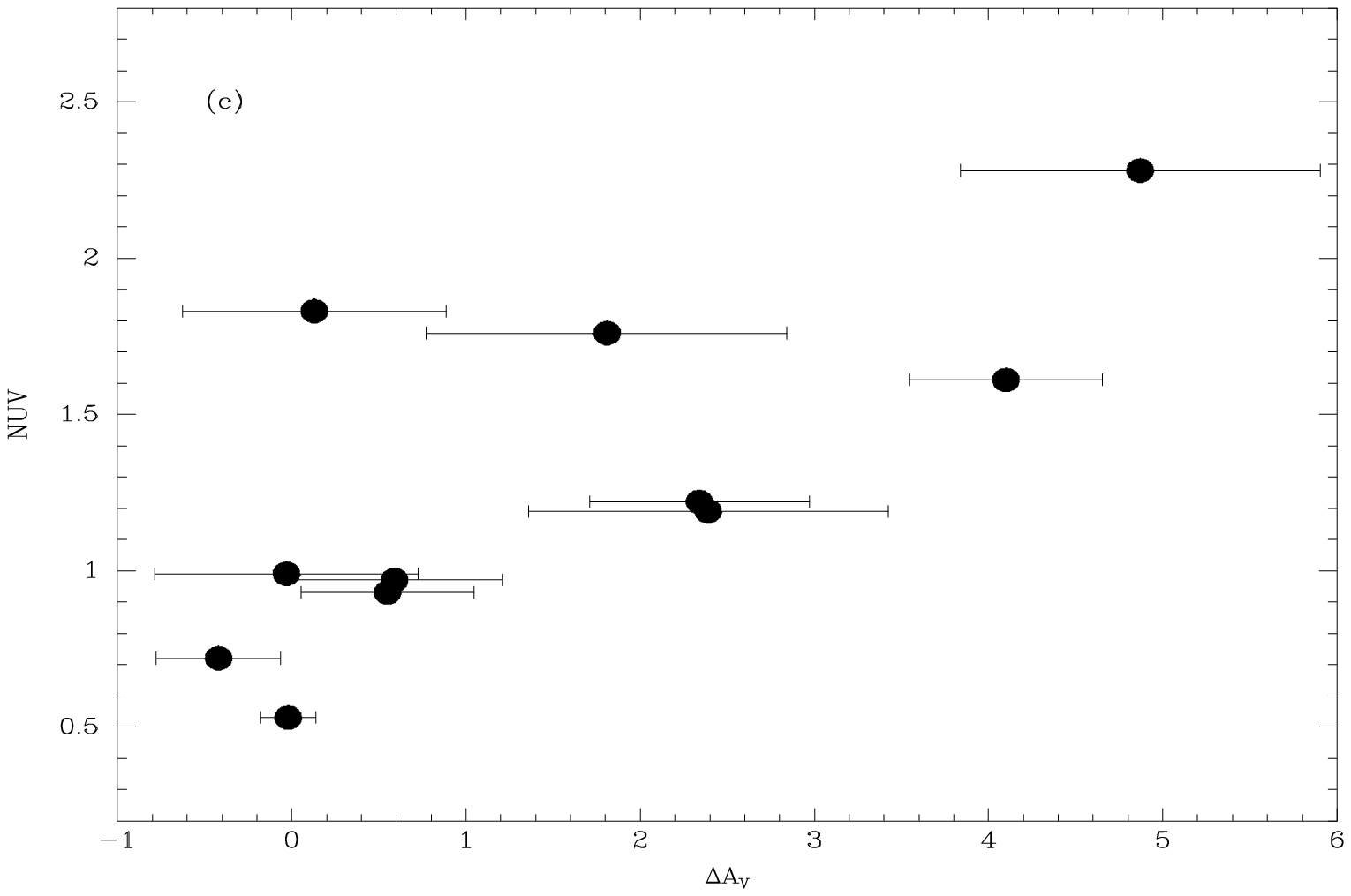}
\plotone{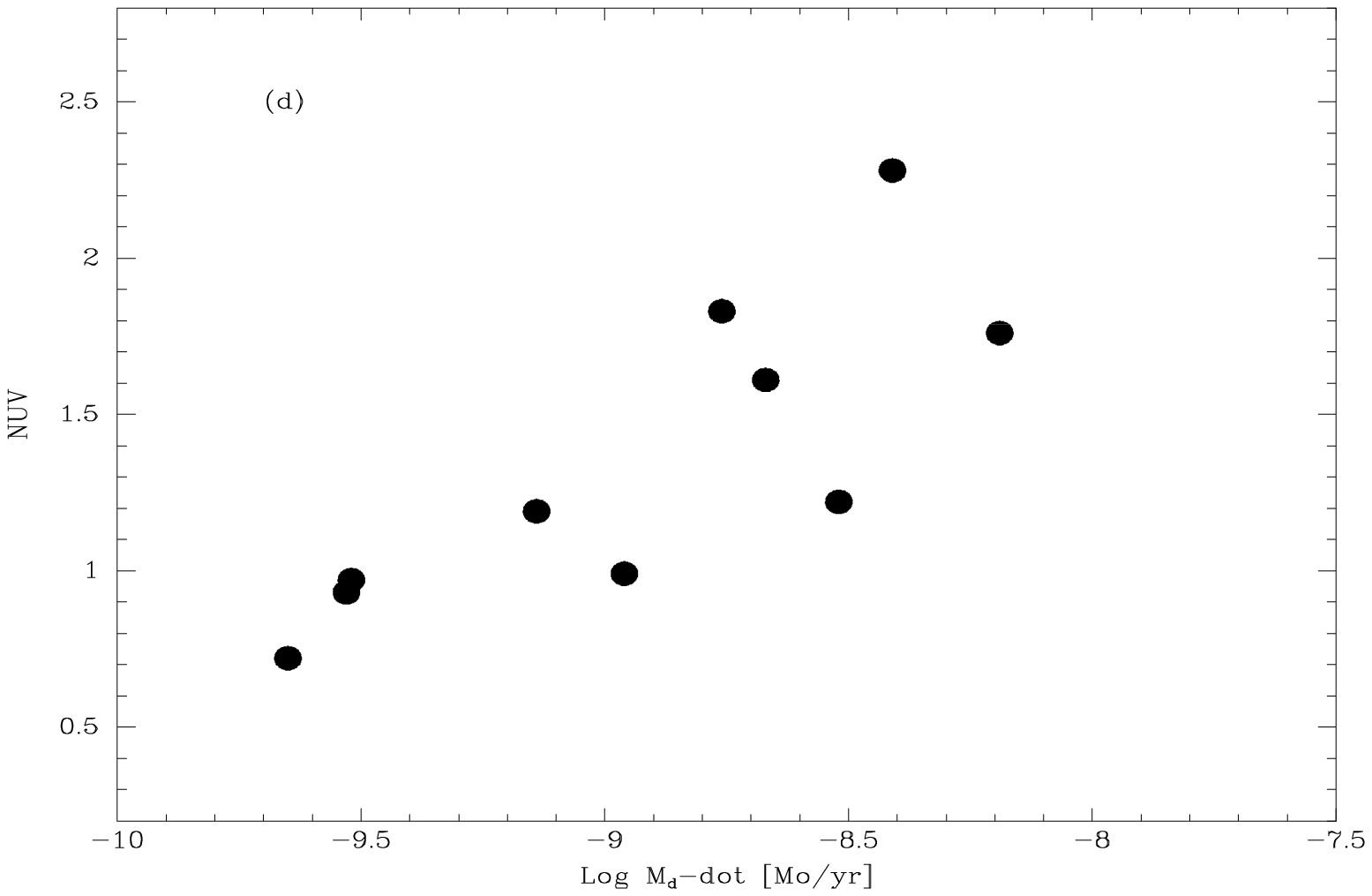}
\plotone{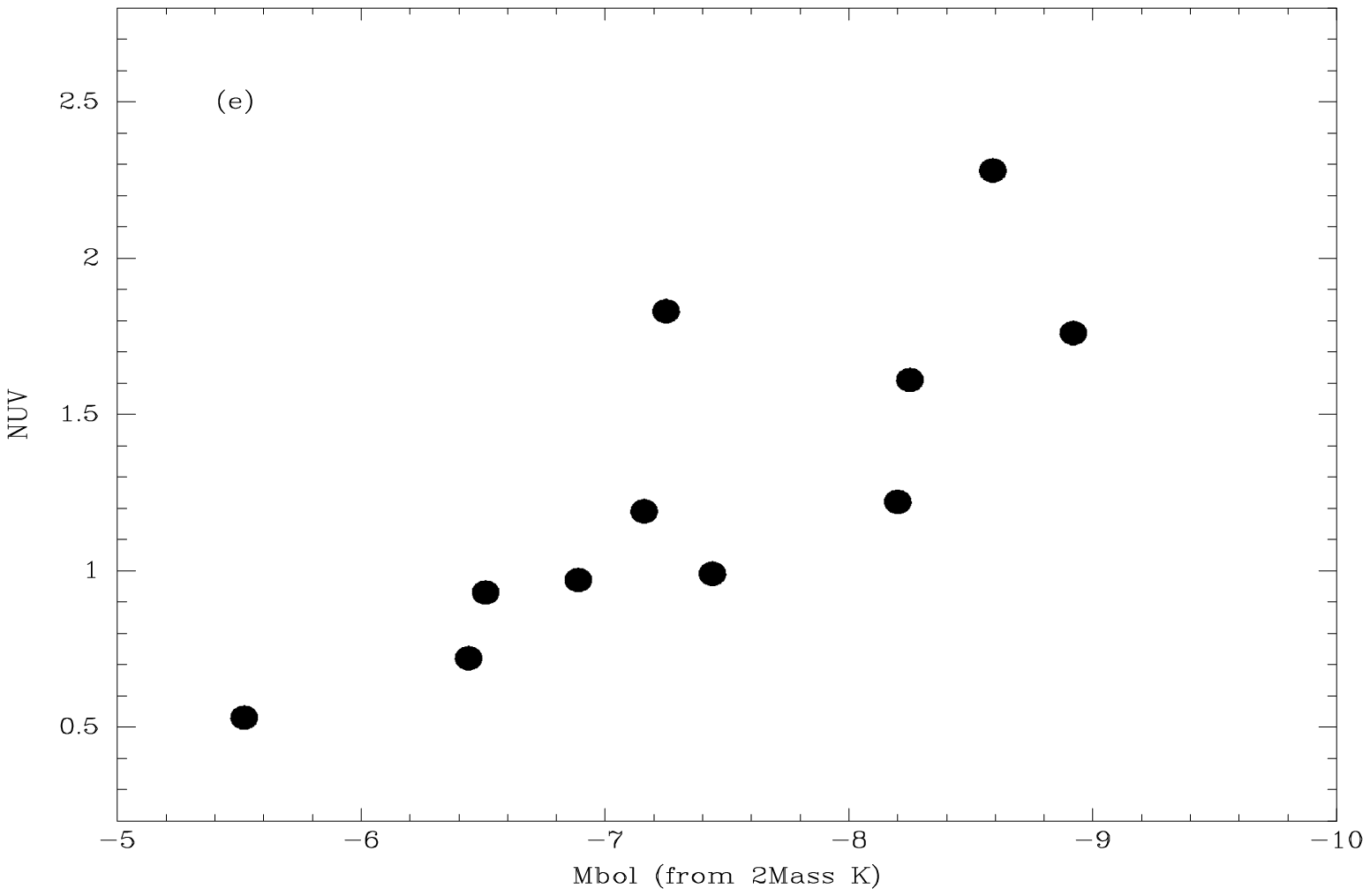}
\caption{\label{fig:nuv} The effect of dust in the NUV. In (a) and (b)
we show the
match between the observations (black solid line) and the reddened
MARCS model (dashed red line) for KY Cyg and HD 216946.  The former
shows considerable extra flux in the NUV compared to the reddened
model; the latter does not. In (c), (d), and (e) we show
the correlation of the NUV index to the extra extinction $\Delta A_V$,
the dust production rate $\dot{M}_d$, and the bolometric luminosity
computed from the K-band, respectively.
}
\end{figure}

In Fig.~\ref{fig:nuv}(a) and (b)
we show the match between the models reddened
by the standard CCM89 $R_V=3.1$ law and the observed spectra for two of
the stars in our sample.  KY~Cyg has the 
the highest excess reddening ($\Delta
A_V=4.9$~mag), and is among the most luminous star in our sample
($M_{\rm bol}=-8.8$).  
The discrepancy in the NUV is striking.
By contrast, the agreement in the NUV region is very good for the
star HD~216946, a star with no excess reddening ($\Delta A_V=0.0$) and
modest luminosity ($M_{\rm bol}=-5.5$).
We can quantify this 
by computing an NUV index where we compare the integrated flux from 3500 to 3900\AA\ 
of the reddened model to that of the star, expressed as a difference in magnitudes.  In the other panels of Fig.~\ref{fig:nuv} we show that the
size of the NUV discrepancy is well correlated with the amount of 
extra extinction (Fig.~\ref{fig:nuv}{\it c}), the dust production rate $\dot{M}_d$ as measured
from the 12-$\mu$m excess (Fig.~\ref{fig:nuv}{\it d}), and the bolometric luminosity of the
star computed from the K-band (Fig.~\ref{fig:nuv}{\it e}).

There are certainly other explanations for the NUV discrepancy.  The
spectra of some stars could be contaminated by the presence of a hot
companion.  However, we would expect such stars to show a composite
spectrum.  Indeed, in Paper~I we had eliminated several stars from our
sample because Balmer lines were clearly present. Another intriguing
possibility is that hot spots might be present on the surface of these
stars, as suggested by interferometric observations and UV {\it HST}
imaging; see Freytag (2003) and references therein. Stars with vigorous
convection patterns could have spectra dominated by regions at
different temperatures depending upon the wavelength. Alternatively,
chromospheric emission may play a role (Carpenter et al.\ 1994,
Harper et al.\ 2001). These possibilities
could, and should, be investigated by higher spectral resolution
studies.  Nevertheless, the correlations shown in Fig.~\ref{fig:nuv}
provide strong evidence that circumstellar dust is the culprit.

Although it is tempting to derive the reddening law of this circumstellar
material, it is difficult to accurately extract the circumstellar
component from the total extinction, given the high (and uncertain)
amount of foreground reddening.  We are in the process of obtaining the
necessary data for a large sample of RSGs in the Magellanic Clouds.  There
the problem is actually tractable, given the small and relatively uniform 
extinction of the Clouds (van den Bergh 2000).

\section{Contribution to the Dust Content of the Milky Way}
\label{Sec-cosmic}

In reviews of the origin of cosmic dust, the role of RSGs is often
ignored, with the primary sources given as SNe and low-mass AGB stars.
However, for dusty galaxies at large look-back times, AGBs cannot play
a role given the time scales for low-mass stellar evolution.  When
high-mass stars are discussed, 
it is usually only
the WC-type Wolf-Rayet stars that are considered (see, for example,
Dwek 1998).
However,  only the late-type WCs are known to produce
dust, and even those types may require a binary companion (see 
Crowther  2003).   In the Milky Way, late-type WCs are concentrated towards
the Galactic center, and are known to be absent in low-metallicity galaxies
(see Massey 2003 and references therein).
So, in extreme environments (such as metal-poor starbursts, or for most
galaxies at early times), RSGs could play a dominant role.

\begin{figure}
\epsscale{0.42}
\plotone{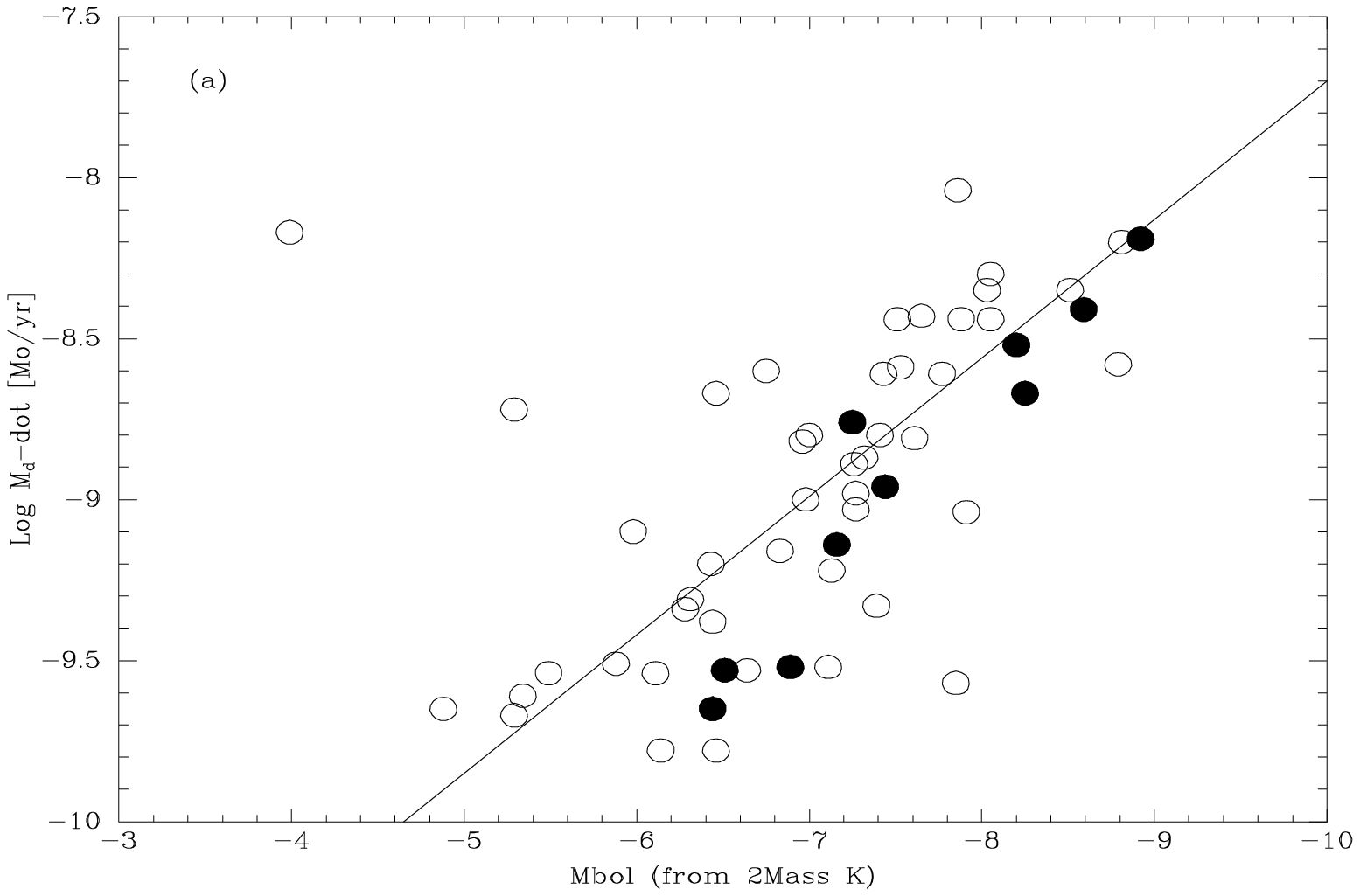}
\plotone{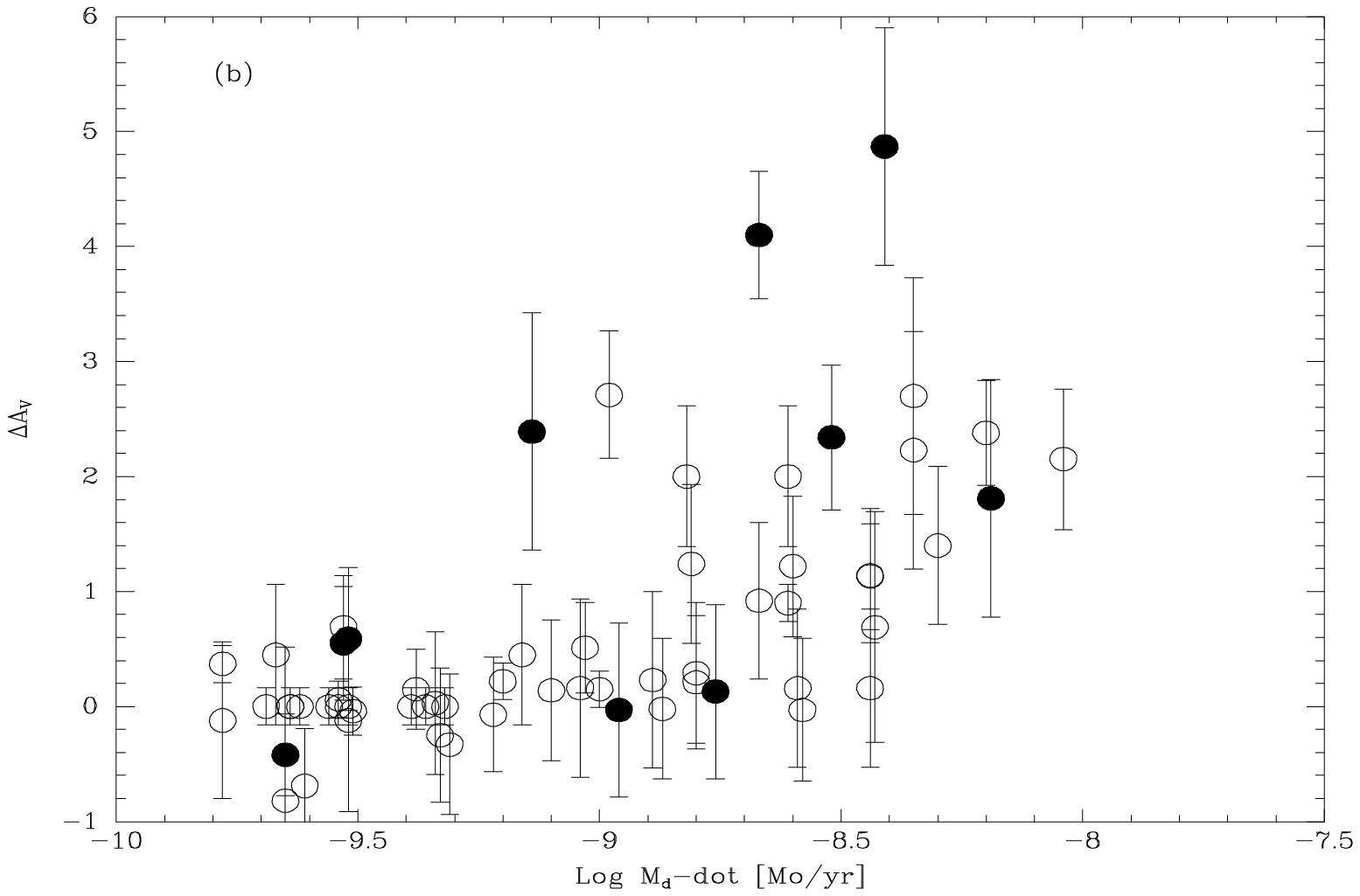}
\plotone{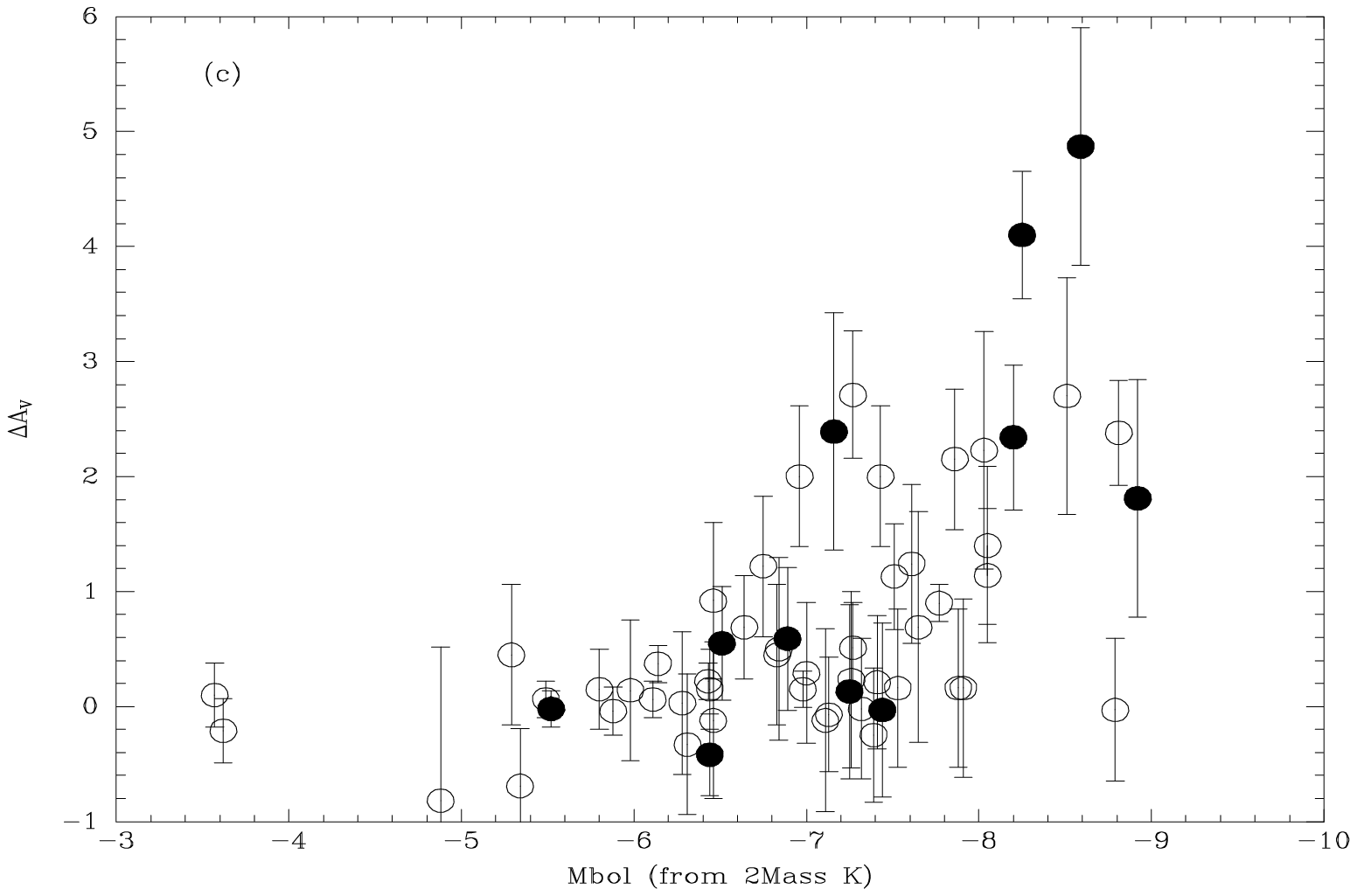}
\caption{\label{fig:eric} The correlation of excess reddening with dust.
In (a) we show that the dust
production rate $\dot{M}_d$ (revised from Josselin et al.\ 2000)
is correlated
with bolometric luminosity. The solid line is the least-squares
fit $\log \dot{M}_d [M_\odot$ yr$^{-1}]=-0.43M_{\rm bol}-12.0$ excluding
four outliers, which are not considered further.
In (b) we compare $\dot{M}_d$ with the excess extinction $\Delta A_V$.
Stars with low dust production rates have extinction that is more in keeping
with that of the early-type stars in the same OB associations; stars with
higher dust production rates also tend to show excess extinction.
In (c) we compare the excess extinction
$\Delta A_V$ to the bolometric luminosity determined from $K$.
In all three plots the filled circles indicate the stars for which we
have new data in the near-UV.}
\end{figure}

Here we briefly consider how large that role is locally.  We expect the production rate
(in units of mass per unit time per unit area of the galactic disk) to be
$$R_{\rm RSG}=\int_{10}^{25} \rho_{\rm RSG} (m) \ \dot{M}_d(m)\  dm,$$
where $\rho_{\rm RSG}(m)$ is the surface density of RSGs as a function of mass and
$\dot{M}_d(m)$ is the dust production rate. 
We can estimate these quantities as follows.
First, from Fig.~\ref{fig:eric}(a), we find that $\log \dot{M}_d=-0.43\times M_{\rm bol} -12.0$
for $M_{\rm bol}<-5$, which roughly corresponds to masses $>10
M_\odot$.  In Paper~I we used the evolutionary tracks of Meynet \&
Maeder (2003) to estimate that the masses of RSGs scale with luminosity
as $\log (M/M_\odot) = 0.50-0.099M_{\rm bol}$,  so we 
expect $$\log \dot{M}_d = 4.3 \log (M/M_\odot) - 14.2$$ for RSGs with
masses $10<(M/M_\odot)<25$.    The 
surface density $\rho_{\rm RSG}(m)$ is harder to estimate. 
We expect $\rho_{\rm RSG} (m) \propto m^{-2.35} \Delta \tau_{\rm RSG} (m)$, where
we have assumed a Salpeter initial mass function, and 
$\Delta \tau_{\rm RSG}$ is the lifetime of the RSG phase as a function of mass.
From the evolutionary models of 
Meynet \& Maeder (2003),  we find that the RSG phase lasts 
2~Myr (10$M_\odot$) to 0.4~Myr (25$M_\odot$), and we use the models
to approximate
$\log \Delta \tau_{\rm RSG}=8.1-1.8\log (M/M_\odot)$.  
Jura \& Kleinmann (1990)  find 21 {\it high-luminosity} ($M_{\rm bol}<-7.8$, corresponding roughly to 19$M_\odot$) RSGs
within 2.5 kpc of the sun, or a surface density of
1 kpc$^{-2}$. If that is complete (and we list a comparable number, 18, 
meeting these criteria
in our admittedly 
incomplete sample from Paper~I), then we can use that to determine
the scaling factor $C$:
$\int_{19}^{25} \rho_{\rm RSG}(m)\  dm= 1.3\times 10^{8} C \int_{19}^{25}  m^{-2.35}\ m^{-1.8}\ dm = 1$, or $C=4.5 \times 10^{-4}$.
Thus $\rho_{\rm RSG}(m) = 5.8\times 10^{4} m^{-4.15}$.
Substituting in the above, we expect 
$$R_{\rm RSG}=3.7\times 10^{-10}  \ \int_{10}^{25}\ m ^{0.15}\ dm,$$
or $8.5 \ 10^{-9} M_\odot$ yr$^{-1}$  kpc$^{-2}$. 
  
We emphasize that this value is uncertain due primarily to our poor
understanding of the exact completeness limits.
We can instead derive the surface density of RSGs using what we do know
about the surface densities of other types of massive stars in the solar
neighborhood, appealing to the evolutionary models to then connect
the two.  Although the number of O-type stars within a few kpc of the
sun is poorly known (see Massey 2003), 
the number of WC-type Wolf-Rayet stars {\it is} thought
to be complete, given their strong emission-line signature
(Massey \& Johnson 1998). According to the
evolutionary models of Meynet \& Maeder (2003), WCs come from stars with
masses greater than 40$M_\odot$; the lifetime of the WC stage is
independent of mass, with $\Delta \tau_{\rm WC}= 0.2$ Myr (see their Fig.\ 10).
The surface density of WCs in the solar neighborhood is 2.1 kpc$^{-1}$,
a number which we believe {\it is} based on a complete sample
(Massey 2003 and references therein).  
So, $\int_{40}^{120}\rho_{\rm WC} (m)\ dm = 2\times 10^{5} C \int_{40}^{120} m^{-2.35}\ dm= 2.1$, and we derive $C=2.7\times 10^{-3}$, about a factor of
6 times greater.  Thus
$R_{\rm RSG}=5.1 \times 10^{-8} M_\odot$ yr$^{-1}$  kpc$^{-2}$. Given the
uncertainties (both in our knowledge of the local RSG content, and in
the models) we consider this agreement very good, and adopt a value
$R_{RSG}=3\times 10^{-8} M_\odot$ yr$^{-1}$  kpc$^{-2}$.  
For comparison Whittet (2003)
estimates a return rate for RSGs that is about 3 time greater, well within
the uncertainties of our approximation.  This is about 1\% of the rate
of return of AGBs.

How does this compare to the production rate for WC-type WRs? 
As stated above, only the late-type WCs (WCLs) are
known
to produce dust, and even those may require a binary companion.  From
Conti \& Vacca (1990) we find a surface density of WCLs of 0.5 kpc$^{-1}$.
The (total) mass-loss rates of WCs are about independent of mass, and
are about $10^{-4.8} M_\odot$ yr$^{-1}$ (Nugis \& Lammers 2000). Of this,
perhaps 2\% is dust (Dwek 1998).  So we expect 
$R_{\rm WCL}=1.5 \times 10^{-7} M_\odot$ yr$^{-1}$  kpc$^{-2}$.  If close binarity
is a requirement for dust ejection, then this value should be decreased by
roughly a factor of 2.  We 
adopt a compromise of $R_{\rm WCL}=1\times 10^{-7} M_\odot$ yr$^{-1}$ kpc$^{-2}$, 
or about a factor of 3 higher than $R_{\rm RSG}$.
In low-metallicity systems (or the outwards of the solar circle
in the Milky Way) there are no late-type WCs, and RSGs will dominate the
dust production by massive stars.  In star-bursts, where stars are recently
formed, most dust production should be by RSGs, as low-mass stars will not
have had sufficient time to evolve to AGBs.

\acknowledgements

We are grateful to Richard Green, the
Director of Kitt Peak National Observatory, for providing 
discretionary time on the 2.1-m telescope to resolve the NUV problem,
and to Di Harmer for support of these efforts.  Conversations and suggestions
by our colleagues, particularly Georges Meynet and Andre Maeder, 
are acknowledged. An anonymous referee was responsible for valuable suggestions,
 including calling our attention to the fact that circumstellar dust should
behave differently in its scattering properties than dust far from the star.

\end{document}